\documentclass[conference]{IEEEtran}

\usepackage{cite}
\usepackage{graphicx}
\usepackage{epsfig,psfrag}
\usepackage{balance}
\usepackage{array}
\usepackage{theorem}
\usepackage{amsmath}
\usepackage{amssymb}
\usepackage{flushend}
\usepackage{tikz}
\usepackage{algorithmic,algorithm}

\newtheorem{theorem}{Theorem}
\newtheorem{fact}[theorem]{Fact}

\newtheorem{proposition}[theorem]{Proposition}
\newtheorem{corollary}[theorem]{Corollary}

\newcommand{\mc}{\mathcal}
\newcommand{\mb}{\mathbf}

\newcommand{\NM}{Ucomp}
\newcommand{\CM}{UcompM}
\newcommand{\UCM}{UcompCM}
\newcommand{\MM}{m}

\newcommand{\TM}{\textup{M}}
\newcommand{\TCM}{\textup{CM}}

\begin{document}
%
\title{Results on the Fundamental Gain of Memory-Assisted Universal Source Coding}
\author{Ahmad Beirami, Mohsen Sardari, Faramarz Fekri\\School of Electrical and Computer Engineering, Georgia Institute of Technology,~Atlanta~GA~30332, USA\\Email:~\{beirami,~mohsen.sardari,~fekri\}@ece.gatech.edu

\thanks{This material is based upon work supported by the National Science Foundation under Grant No. CNS-1017234.}
}
%
%
%

\maketitle
\thispagestyle{empty}
\pagestyle{empty}

\begin{abstract}
Many applications require data processing to be performed on individual pieces of data which are of finite sizes, e.g., files in cloud storage units and packets in data networks. However, traditional universal compression solutions would not perform well over the finite-length sequences. Recently, we proposed a framework called memory-assisted universal compression that holds a significant promise for reducing the amount of redundant data from the finite-length sequences. The proposed compression scheme is based on the observation that it is possible to learn source statistics (by memorizing previous sequences from the source) at some intermediate entities and then leverage the memorized context to reduce redundancy of the universal compression of finite-length sequences.
We first present the fundamental gain of the proposed memory-assisted universal source coding over conventional universal compression (without memorization) for a single parametric source.
Then, we extend and investigate the benefits of the memory-assisted universal source coding when the data sequences are generated by a compound source which is a mixture of parametric sources.
We further develop a clustering technique within the memory-assisted compression framework to better utilize the memory by classifying the observed data sequences from a mixture of parametric sources.
Finally, we demonstrate through computer simulations that the proposed joint memorization and clustering technique can achieve up to 6-fold improvement over the traditional universal compression technique when a mixture of non-binary Markov sources is considered.
\end{abstract}


\section{Introduction}
\label{sec:introduction}
\vspace{-0.03in}
Since Shannon's seminal work on the analysis of communication systems, many researchers have contributed  toward the development of compression schemes with the average code length as close as possible to the entropy.
In practice, we usually cannot assume a priori knowledge on the statistics of the source although we still wish to compress the \emph{unknown} stationary ergodic source to its entropy rate. This is known as the \emph{universal} compression problem~\cite{Davisson_noiseless_coding, Universal_Finite_memory_source, CTW95}.
However, unfortunately, universality imposes an inevitable redundancy depending on the richness of the class of the sources with respect to which the code is universal~\cite{Merhav_Feder_IT,ISIT11,Rissanen_1984}.
While an entire library of concatenated sequences from the same context (i.e., source model) can usually be encoded to less than a tenth of the original size using universal compression~\cite{ISIT11,Rissanen_1984},
it is usually not an option to concatenate and compress the entire library at once.
On the other hand, when an individual sequence  is universally compressed regardless of other sequences, the performance is fundamentally limited~\cite{Merhav_Feder_IT,ISIT11}.

In~\cite{Korodi2005},  the authors observed that forming a statistical model
from a training data would improve the performance
of universal compression on finite-length sequences.  In~\cite{DCC12}, we introduced \emph{memory-assisted universal source coding}, where we proposed memorization of the previously seen sequences as a
solution that can fundamentally improve the performance of universal compression.
As an application of memory-assisted compression, we introduced the notion of \emph{network compression} in~\cite{ITW11, INFOCOM12}. It was shown that by deploying memory in the network (i.e., enabling some nodes to memorize source sequences), we may remove the redundancy in the network traffic.
In~\cite{ITW11,INFOCOM12}, we assumed that memorization of the previous sequences from the same source provides a fundamental gain $g$ over and above the conventional compression performance of the universal compression of a new sequence from the same source. Given $g$, we derived the network-wide memorization gain $\mc{G}$ on both a random network graph~\cite{ITW11} and a power-law network graph~\cite{INFOCOM12} when a small fraction of the nodes in the network are capable of memorization. However,~\cite{ITW11, INFOCOM12} did not explain as to how $g$ is computed.

\begin{figure}
\centering
\includegraphics[height=1.55in, angle=-90]{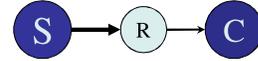}
\caption{Memory-assisted compression in a two-hop communication scenario.} \vspace{-0.05in}
\label{fig:single_hop}
\vspace{-0.2in}
\end{figure}

Although the memory-assisted universal source coding naturally arises in a various set of problems, we define the problem setup in the most basic network scenario depicted in Fig~\ref{fig:single_hop}.
We assume that the network consists of the server  $S$, the intermediate (relay) node $R$, and the client $C$, where $S$ wishes to send the sequence $x^n$ to $C$. We assume that $C$ does not have any prior communication with the server, and hence, is not capable of memorization of the source context.  However, as an intermediate node, $R$ has observed several previous sequences from $S$ when forwarding them from $S$ to clients other than $C$ (not shown in Fig.~\ref{fig:single_hop}). Therefore, $R$ has formed a memory of the previous communications shared with $S$.
Note that if the intermediate node $R$ was absent, the source could possibly apply universal compression to $x^n$ and transmit to $C$ whereas the presence of the  memorized sequences at $R$ can potentially reduce the communication overhead in the $S$-$R$ link.

The objective of the present paper is to characterize the fundamental gain $g$ of memorization of the context from a server's previous sequences in the universal compression of a new individual sequence from the same server.
 Clearly, a single stationary ergodic source does not fully model a real content generator server (for example the CNN news website in the Internet). Instead, a better model is to view every content generator server as a compound (mixture) of several information sources whose true statistical models are not readily available. In this work, we try to address this issue and propose a memorization and clustering technique for compression that is suitable for a compound source. Namely, we would like to answer the following questions in the above setup:
1)  Would the deployment of memory in the encoder ($S$) and the decoder ($R$) provide any fundamental benefit in the universal compression?
2) If so, how does this gain $g$ vary as the sequence length $n$ and the memorized context length $m$ change?
3) How much performance improvement should we expect from the joint memorization and clustering versus the memorization without clustering?
4) How should we realize the clustering scheme to achieve good performance from compression with the joint memorization and clustering?

\section{Background Review and Motivation}
\label{sec:background}
In this section, we motivate the context memorization problem by demonstrating the significance of redundancy in the universal compression of small to moderate length sequences.
Let $\mc{A}$ be a finite alphabet. Let the parametric source be defined using a $d$-dimensional parameter vector $\theta = (\theta_1,...,\theta_d)$, where $d$ denotes the number of the source parameters. Denote $\mu_\theta$ as the probability measure defined by the parameter vector $\theta$ on sequences of length $n$. We also use the notation $\mu_\theta$ to refer to the parametric source itself. We assume that the $d$ parameters are unknown.
Denote $\mc{P}^d$ as the \emph{family} of sources with $d$-dimensional unknown parameter vector $\theta$. We use the notation $x^n = (x_1,...,x_n) \in \mc{A}^n$ to present a sequence of length $n$ from the alphabet $\mc{A}$.

 Let $H_n(\theta)$ be the source entropy given $\theta$, i.e.,
\begin{equation}
H_n(\theta) \hspace{-.02in}= \hspace{-.02in} \mb{E} \log \hspace{-.03in}\left(\frac{1}{\mu_\theta(X^n)} \right)\hspace{-.04in} =\hspace{-.02in}\sum_{x^n}\mu_{\theta}(x^n) \log \hspace{-.03in}\left(\frac{1}{\mu_\theta(x^n)}\right)\hspace{-.03in}.\footnote{Throughout this paper expectations are taken over the random sequence $X^n$ with respect to the probability measure $\mu_\theta$ unless otherwise stated.}
\end{equation}
In this paper $\log(\cdot)$ always denotes the logarithm in base $2$.
 Let $c_n:\mc{A}^n \to \{0,1\}^*$ be an injective mapping from the set $\mc{A}^n$ of the sequences of length $n$ over $\mc{A}$ to the set $\{0,1\}^*$ of binary sequences. Further, let $l_n(x^n)$ denote a universal length function for the codeword associated with the sequence $x^n$.
Denote $R_n(l_n,\theta)$ as the expected redundancy of the code with length function $l_n(\cdot)$, defined as $R_n(l_n,\theta) = \mb{E}l_n(X^n) -  H_n(\theta)$.
Note that the expected redundancy is always non-negative. 

Let $\mc{I}_n(\theta)$ be the Fisher information matrix, i.e.,
\begin{equation}
\mc{I}_n(\theta)\hspace{-.03in} =\hspace{-0.01in} \{\mc{I}_n^{ij}(\theta)\} \hspace{-.03in}= \hspace{-.01in}\frac{1}{n\log e}\mb{E}\hspace{-.03in}\left\{\frac{\partial^2}{\partial \theta_i \partial \theta_j} \log \hspace{-.03in}\left( \frac{1}{\mu_\theta(X^n)}\right)\hspace{-.03in} \right\}\hspace{-.02in}.
\end{equation}
Fisher information matrix quantifies the amount of information, on the average, that each symbol in a sample sequence $x^n$ from the source conveys about the source parameters.
Let Jeffreys' prior on the parameter vector $\theta$ be denoted by $\omega_J(\theta) \triangleq \frac{|\mc{I}(\theta)|^{\frac{1}{2}}}{\int|\mc{I}(\lambda)|^\frac{1}{2}d\lambda}$.
Jeffreys' prior is optimal in the sense that the average minimax redundancy is achieved when the parameter vector $\theta$ is assumed to follow Jeffreys' prior~\cite{Clarke_Barron}.
Further, let $\bar{R}_n$ be the average minimax redundancy given by~\cite{Clarke_Barron,atteson_markov}
\begin{equation}
\bar{R}_n = \frac{d}{2}  \log\left( \frac{n}{2\pi e} \right) + \log \int
|\mc{I}_n(\theta)|^{\frac{1}{2}}d\theta + O\left(\frac{1}{n}\right).
\label{eq:minimax}
\end{equation}

In~\cite{ISIT11}, we obtained a lower bound on the average redundancy of the universal compression for the family of conditional two--stage codes, where the unknown parameter is first estimated and the sequence is encoded using the estimated parameter, as the following~\cite{ISIT11}:
\begin{theorem}
Assume that the parameter vector $\theta$ follows Jeffreys' prior in the universal compression of the family of parametric sources $\mc{P}^d$.
Let $\delta$ be a real number. Then,
\begin{equation}
\mb{P}\left[ \frac{R_n(l_n,\theta)}{\frac{d}{2}\log n} \geq 1- \delta \right] \geq 1 - \frac{1}{\int |\mc{I}(\lambda)|^{\frac{1}{2}}d \lambda} \left(\frac{2 \pi e}{n^\delta} \right)^{\frac{d}{2}}.\nonumber
\end{equation}
\label{thm:Main_general}
\end{theorem}
Theorem~\ref{thm:Main_general} can be viewed as a tighter variant of  Theorem~1 of Merhav and Feder in~\cite{Merhav_Feder_IT} for parametric sources.

\begin{figure}[tb]
\centering
\vspace{-0.05in}
\psfrag{xlabel}{\footnotesize{$n$}}
\psfrag{ylabel}{\footnotesize{$\frac{\mb{E}\{l_n(X^n)\}}{H_n(\theta)}$}}
\psfrag{dataaaaaaaaaa1}{\tiny{$H_n(\theta)/n = \frac{1}{2}$}}
\psfrag{data2}{\tiny{$H_n(\theta)/n = 1$}}
\psfrag{data3}{\tiny{$H_n(\theta)/n = 2$}}
\psfrag{data4}{\tiny{$H_n(\theta)/n = 4$}}
\psfrag{256kb}{\footnotesize{256kB}}
\psfrag{1Mb}{\footnotesize{1MB}}
\psfrag{4Mb}{\footnotesize{4MB}}
\psfrag{16Mb}{\footnotesize{16MB}}
\psfrag{64Mb}{\footnotesize{64MB}}
\epsfig{width= 0.8\linewidth,file=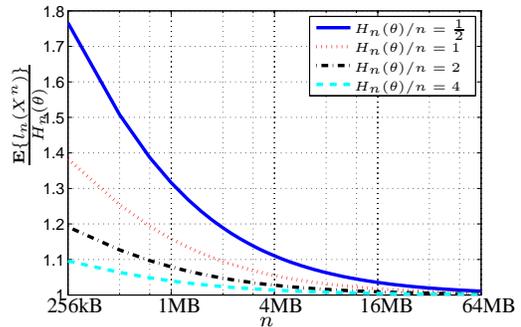}
\vspace{-0.05in}
\caption{The Lower bound on compression for at least $95\%$ of the sources as a function of sequence length~$n$ for different values of entropy rate $H_n(\theta)/n$.}
\vspace{-0.05in}
\label{fig:M1_256_overhead}
\end{figure}

To demonstrate the significance of the above theorem, we consider an example using a first-order Markov source with alphabet size $k=256$. This source may be represented using $d = 256 \times 255 = 62580$ parameters.
Fig.~\ref{fig:M1_256_overhead} shows the average number of bits per symbol required to compress the class of the first-order Markov sources normalized to the entropy of the sequence for different values of entropy rates in bits per source symbol (per byte).
 In this figure, the curves demonstrate the lower bound on the compression rate achievable for at least $95\%$ of sources , i.e., the probability measure of the sources from this class that may be compressed with a redundancy smaller than the curve is at most $\epsilon = 0.05$.
 As can be seen, if the source entropy rate is $1$ bit per byte ($H_n(\theta)/n = 1$), the compression overhead is $38\%$, $16\%$, $5.5\%$, $1.7\%$, and $0.5\%$ for sequences of lengths $256$kB, $1$MB, $4$MB, $16$MB, and $64$MB, respectively. Hence,  we conclude that redundancy is significant in the compression of finite-length low-entropy sequences, such as the Internet traffic. It is this redundancy that we hope to remove using the memorization technique.

\section{Fundamental Gain of Context Memorization}
\label{sec:gain_setup}
In this section, we present the problem setup and define the context memorization gain.
We  assume that the compound source comprises of a mixture of $\mc{K}$ information sources.
Denote $[\mc{K}]$ as the set $\{1,...,\mc{K}\}$. As the first step, in this paper, we assume that $\mc{K}$ is finite and fixed.
We consider parametric sources with $\theta^{(i)}$ as the parameter vector for the source $i$ ($i\in [\mc{K}]$). As in~\cite{ISIT11}, we assume that $\theta^{(i)} = (\theta_1^{(i)}, \theta_2^{(i)},\ldots,\theta_d^{(i)})$ follows Jeffreys' prior for all $i\in [\mc{K}]$.  We consider the following scenario.
We assume that, in Fig.~\ref{fig:single_hop}, both the encoder (at $S$) and the decoder (at $R$) have access to a memory of the previous $T$ sequences from the compound source.
Let $\mb{m} =(n_0,\ldots, n_{T-1})$ denote the lengths of the previous $T$ sequences generated by $S$. Further, denote $\mb{y} = \left\{y^{n_j}(j)\right\}_{j=0}^{T-1}$ as the previous $T$ sequences from $S$ visited by the memory unit $R$.
Note that each of these sequences might be from a different source model. We denote $\mb{p} = (p_1,...,p_\mc{K})$, where $\sum_{i=1}^\mc{K} p_i = 1$, as the probability distribution according to which the information sources in the compound source are selected for sequence generation, i.e., the source $i$ is picked with probability $p_i$.
Let the random variable $Z_j$ denote the index of the source that has generated the sequence $y^{n_j}(j)$, and hence, $Z_j$ follows the distribution $\mb{p}$ over $[\mc{K}]$.
Therefore, at time step $j$, sequence $y^{n_j}(j)$ is generated using the parameter vector $\theta^{(Z_j)}$. Further, denote $\mb{Z}$ as the vector $\mb{Z} = (Z_0,...,Z_{T-1})$.
We wish to compress the sequence $x^n$ with source index $Z_T$, when both the encoder and the decoder have access to a realization $\mb{y}$ of the random vector $\mb{Y}$.
This setup, although very generic, can incur in many applications. As the most basic example, consider the communication scenario in Fig.~\ref{fig:single_hop}. The presence of memory $\mb{y}$ at $R$ can be used by $S$ to compress (via memory-assisted source coding) the sequence $x^n$ which is requested by client $C$ from $S$. The compression can reduce the transmission cost on the $S-R$ link while being transparent to the client, i.e., $R$ decodes the memory-assisted source code and then applies conventional universal compression to $x^n$ and transmits to $C$.

In order to investigate the fundamental gain of the context memorization in the memory-assisted universal compression of the sequence $x^n$ over conventional universal source coding, we compare the following three schemes.
\begin{itemize}
\item {\NM} (Universal compression), in which a sole universal compression is applied on the sequence $x^n$ without regard to the memorized sequence $\mb{y}$.
\item {\CM} (Universal compression with context memorization), in which the encoder $S$ and the decoder $R$ both have access to the memorized sequence $\mb{y}$ from the compound source, and they use $\mb{y}$ to learn the statistics of the source for the  compression of the sequence $x^n$.
\item {\UCM} (Universal compression with source-defined clustering of the memory), which assumes that the memory $\mb{y}$ is shared between the encoder $S$ and the decoder $R$ (i.e., the memory unit). Further, the source defined clustering of memory implies that both $S$ and $R$ exactly know the index $\mb{Z}$ of the memorized sequences.
\end{itemize}
The performance of {\NM} is characterized using the expected redundancy $R_n(l_n,\theta)$, which is discussed in Sec.~\ref{sec:background}.
Let $Q(l_n, \hat{l}_n, \theta)$ be defined as the ratio of the expected codeword length with length function $l_n$ to that of $\hat{l}_n$, i.e.,
\vspace{-0.08in}
\begin{equation}
Q(l_n, \hat{l}_{n}, \theta) \triangleq \frac{\mb{E} l_n(X^n)}{\mb{E} \hat{l}_{n}(X^n)} = \frac{H_n(\theta) + R_n(l_n,\theta)}{H_n(\theta) + R_n(\hat{l}_n,\theta)}.
\vspace{-0.08in}
\end{equation}
Further, let $\epsilon$ be such that $0<\epsilon<1$.
We define $g(l_n,\hat{l}_n, \epsilon)$ as the gain of the length function $\hat{l}_n$ as compared to $l_n$. That is
\vspace{-0.08in}
\begin{equation}
g(l_n,\hat{l}_n,\theta,\epsilon) = \sup_{z \in \mathbb{R}} \left\{z : \mb{P}\hspace{-0.04in}\left[Q(l_n, \hat{l}_{n}, \theta)\geq z\right] \geq 1-\epsilon\right\}.
\label{eq:g_definition}
\vspace{-0.07in}
\end{equation}

In the case of {\CM}, let $l_{n|\mb{m}}$ be the length function with context memorization, where the encoder $S$ and the decoder $R$ have access to sequences $\mb{y}$ with lengths $\mb{m}$. Let $\MM \triangleq |\mb{m}|= \sum_{j=0}^{T-1} n_j$ denote the total length of memory.\footnote{We assume that $n_j\gg h$, where $h$ is the height of the tree of the class $\mc{P}^d$, and hence, the impact of the concatenation of the sequences is negligible.} Further, let $\phi \triangleq \theta^{(Z_T)}$. Denote $R_n(l_{n|\mb{m}},\phi)$ as the expected redundancy of encoding a sequence of length $n$ form the parametric source $\mu_{\phi}$ using the length function $l_{n|\mb{m}}$.
We denote $g_\TM(n,m,\phi,\epsilon, \mb{p}) \triangleq \mb{E}_\mb{Z} g(l_n,l_{n|\mb{m}},\phi, \epsilon)$ as  the fundamental gain of the context memorization on the family of parametric sources $\mc{P}^d$ on a sequence of length $n$ using context memory lengths $\mb{m}$ for a fraction $(1-\epsilon)$ of the sources.
In other words, context memorization provides a gain at least $g_\TM(n,m,\phi,\epsilon,\mb{p})$ for a fraction $(1-\epsilon)$ of the sources in the family.

Similarly in the case of {\UCM}, let $l_{n|\mb{m}, \mb{Z}}$ denote the length function for the universal compression of a sequence of length $n$ with memorized sequences $\mb{y}$, where the vector $\mb{Z}$  of the source indices is known. We denote $g_{\text{CM}}(n,m,\phi,\epsilon,\mb{p}) \triangleq \mb{E}_\mb{Z} g(l_n,l_{n|\mb{m}, \mb{Z}} , \phi,\epsilon)$ as  the fundamental gain of the context memorization in UcompCM.
The following is a trivial lower bound on the context memorization gain in UcompCM.
\vspace{-0.13in}
\begin{fact}
The fundamental gain of context memorization is: $g_\TCM(n,m,\phi,\epsilon,\mb{p}) \geq 1.$
\label{thm:g_1}
\vspace{-0.06in}
\end{fact}
Fact~\ref{thm:g_1} simply states that the context memorization with source defined clustering does not worsen the performance of the universal compression.
We stress again that the saving of memory-assisted compression in terms of flow reduction is only obtained in the $S$-$R$ link. For example, for the given memorization gain $g_\TCM(n,m,\phi,\epsilon,\mb{p}) = g_0$, the expected number of bits  needed to transfer $x^n$ to $R$ is reduced from $\mb{E}l_n(X^n)$ in {\NM} to $\frac{1}{g_0} \mb{E} l_n(X^n)$ in {\UCM}.


\vspace{-0.07in}
\section{Results on the Memorization Gain}
\vspace{-.03in}
\label{sec:results}
In this section, we present our main results on the memorization gain with and without clustering. The proofs are omitted due to the lack of space. We give further consideration to the case $\mc{K} = 1$ since it represents the memorization gain when  all of the memorized sequences are from a single fixed source model.

\vspace{-0.12in}
\subsection{Case $\mc{K}=1$}
\vspace{-0.04in}
In this case, since $\mb{p} = 1$ and $\mb{Z} = \mb{1}$ is known, there is no distinction between {\CM} and {\UCM}, and hence, we drop the subscript of $g$.
 The next theorem characterizes the fundamental gain of memory-assisted source coding:
\begin{theorem}
Assume that the parameter vector $\theta$ follows Jeffreys' prior in the universal compression of the family of parametric sources $\mc{P}^d$. Then,
\vspace{-0.05in}
\begin{equation}
g(n,m,\phi,\epsilon,1)
\geq 1 + \frac{\bar{R}_n + \log(\epsilon )  - \hat{R}_1(n,\MM)}{H_n(\phi) + \hat{R}_1(n,\MM)}  + O\left(\frac{1}{n\sqrt{\MM}} \right),\nonumber
\end{equation}
where $\hat{R}_1(n,\MM) \triangleq \frac{d}{2}\log\left( 1+ \frac{n}{\MM}\right) + 2$.
\label{thm:Gain_K1}
\end{theorem}
Further, let $g(n,\infty, \phi, \epsilon,\mb{p})$ be defined as the achievable gain of  memorization where there is no constraint on the length of the memory, i.e, $g(n,\infty, \phi,\epsilon,\mb{p}) \triangleq \lim_{\MM \to \infty} g(n,\MM,\phi, \epsilon,\mb{p})$. The following Corollary quantifies the  memorization gain for unbounded memory size.
\begin{corollary}
Assume that the parameter vector $\theta$ follows Jeffreys' prior in the universal compression of the family of parametric sources $\mc{P}^d$. Then,
\begin{equation}
g(n,\infty,\phi, \epsilon,1 )
\geq 1 + \frac{\bar{R}_n + \log(\epsilon )  - 2}{H_n(\phi) + 2}.\nonumber
\end{equation}
\label{thm:gain_M}
\end{corollary}
Next, we consider the case where the sequence length $n$ grows to infinity. Intuitively, we would expect that the  memorization gain become negligible for the compression of long sequences.
Let $g(\infty,\MM, \phi, \epsilon,\mb{p}) \triangleq \lim_{n \to \infty} g(n,\MM,\phi,\epsilon,\mb{p})$.
In the following, we claim that memorization does not provide any benefit  when $n\to \infty$:
\begin{proposition}
$g(n,\MM, \phi,\epsilon,1)$ approaches $1$ as the length of the sequence $x^n$ grows, i.e., $g(\infty,\MM, \phi,\epsilon,1)=1$.
\label{thm:Gain_infinite_n}
\end{proposition}

\subsection{{\CM}: Case $\mc{K} \geq 2$}
\label{subsec:gain_UcompM}
As stated in the problem setup, the sequences in the memory may be from various sources. This raises the question that whether a naive memorization of the previous sequences using {\CM} without regard to which source parameter has indeed generated the sequence would suffice to achieve the memorization gain.
Let $\mc{D}$ be an upper bound on the size of the context tree used in compression. Denote $\bar{\mu}_\theta^\mc{D}$ as the probability measure that is defined on the tree of depth $\mc{D}$ from the mixture of the sources.
Further, let $D_n(\mu_\phi||\mu_{\bar{\theta}}) = \sum_{x^n} \mu_\phi(x^n) \log \left(\frac{\mu_\phi(x^n)}{\mu_{\bar{\theta}}(x^n)}\right)$.
The following proposition characterizes the performance of UcompM when applied to a compound source for $\mc{K}\geq 2$.
\begin{proposition}
Let $\theta^{(i)}$ ($i \in [\mc{K}]$) follow Jeffreys' prior.
Then, the memorization gain in UcompM as $m \to \infty$ is upper bounded by
\begin{equation}
 g_\TM(n,\infty,\phi,\epsilon,\mb{p}) \leq \frac{H_n(\phi) + \bar{R}_n}{H_n(\phi)+ D_n(\mu_\phi || \bar{\mu}_{\theta}^\mc{D} )} + O\left(\frac{1}{n}\right).\nonumber
\end{equation}
\label{lem:no_cluster}
\end{proposition}
Note that since $\mc{K}\geq 2$, then $D_n(\mu_\phi || \bar{\mu}_{{\theta}}^\mc{D}) = \Theta(n)$,\footnote{$f(n) = \Theta(g(n))$ if and only if $f(n) = O(g(n)$ and $g(n) = O(f(n))$.}(unless $\theta^{(i)} = \theta^{(j)}$ for all $i,j \in [\mc{K}]$, which occurs with zero probability).
Therefore, the redundancy of UcompM is $R_n(l_{n|\mb{m}}, \phi) = \Theta(n)$ with probability one.
Proposition~\ref{lem:no_cluster} states that when the context is built using the mixture of the sources,  with probability one, the redundancy of UcompM is worse than the redundancy of Ucomp for a sufficiently large sequence, i.e., the memorization gain becomes less than unity for sufficiently large $n$. Therefore, the crude memorization of the context by node $R$ in Fig.~\ref{fig:single_hop} from the previous communications not only does not improve the compression performance but also asymptotically makes it worse. We shall see some discussion on validation of this claim based on simulations in Sec.~\ref{sec:simulation}.

\begin{figure*}[tbh]
\begin{minipage}[b]{0.3\textwidth}
 \psfrag{xlabel}{\footnotesize{$n$}}
\psfrag{ylabel}{\hspace{-.3in}\footnotesize{$g(n,\MM,\phi,0.05,1)$}}
\psfrag{datainf}{\tiny{$\MM \to \infty$}}
\psfrag{dataaaaaaa1}{\tiny{$\MM =$ 8MB}}
\psfrag{data2}{\tiny{$\MM =$ 2MB}}
\psfrag{data3}{\tiny{$\MM =$ 512kB}}
\psfrag{data4}{\tiny{$\MM =$ 128kB}}
\psfrag{256kb}{\tiny{128kB}}
\psfrag{1Mb}{\tiny{512kB}}
\psfrag{4Mb}{\tiny{2MB}}
\psfrag{16Mb}{\tiny{8MB}}
\psfrag{64Mb}{\tiny{32MB}}
\epsfig{width= 1.05\linewidth, height=0.8\linewidth, file=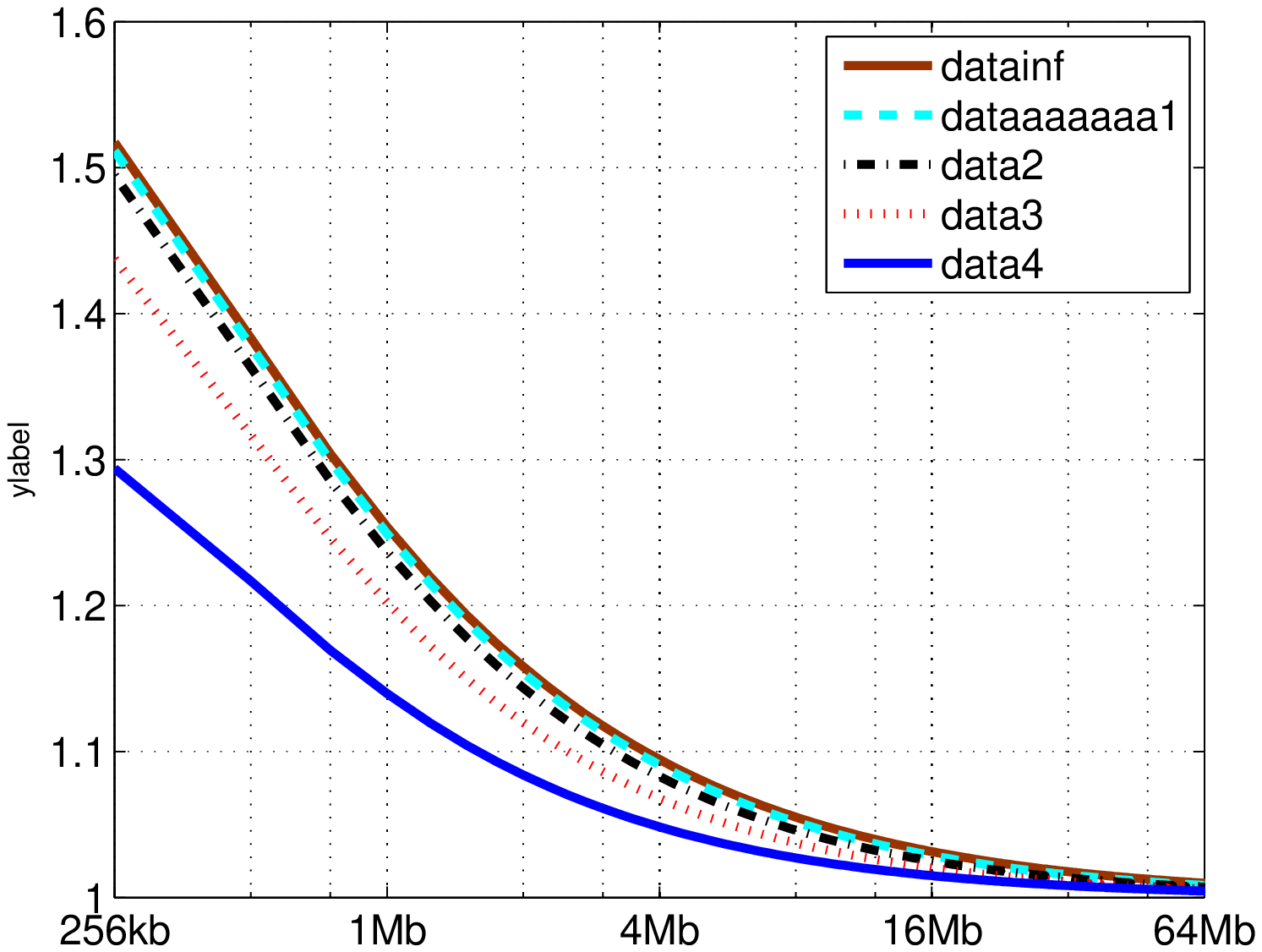}
\vspace{-0.25in}
\caption{Theoretical lower bound on the memorization gain~$g(n,\MM,\phi,0.05,1)$.}
 \vspace{-0.1in}
\label{fig:M1_256_gain_n}
  \end{minipage}
\hspace{0.03\textwidth}
  \begin{minipage}[b]{0.3\textwidth}
  \centering
  \vspace{-0.1in}
  \includegraphics[width=1.11\linewidth, height=0.8\linewidth]{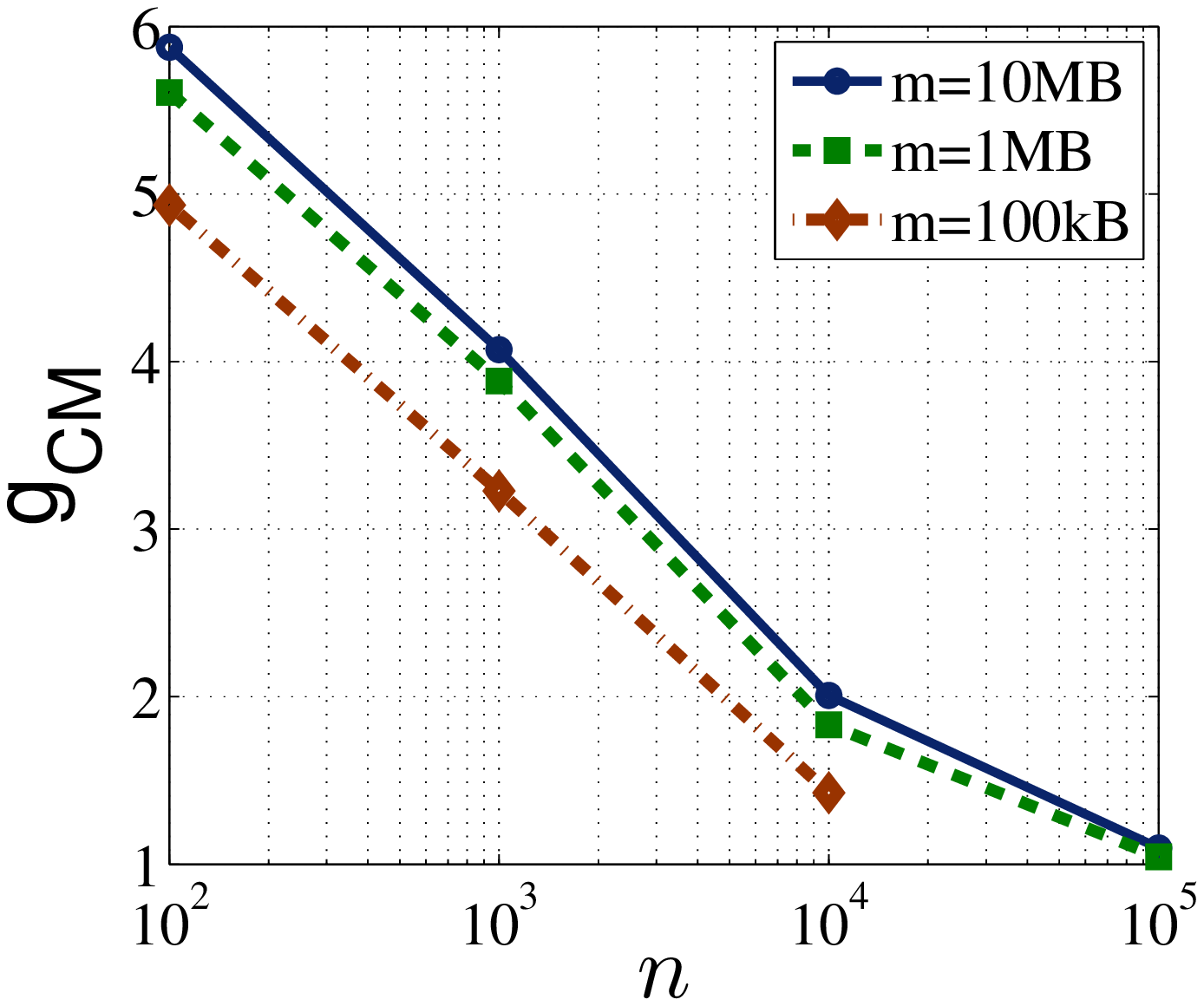}
  \vspace{-0.27in}
  \caption{The gain $g_\text{CM}$ of memory-assisted compression with source-defined clustering.}
  \label{fig:g_pc}
  \end{minipage}
\hspace{0.03\textwidth}
  \begin{minipage}[b]{0.3\textwidth}
  \centering
  \vspace{-0.4in}
  \includegraphics[height=1.11\linewidth, width=0.8\linewidth,angle=-90]{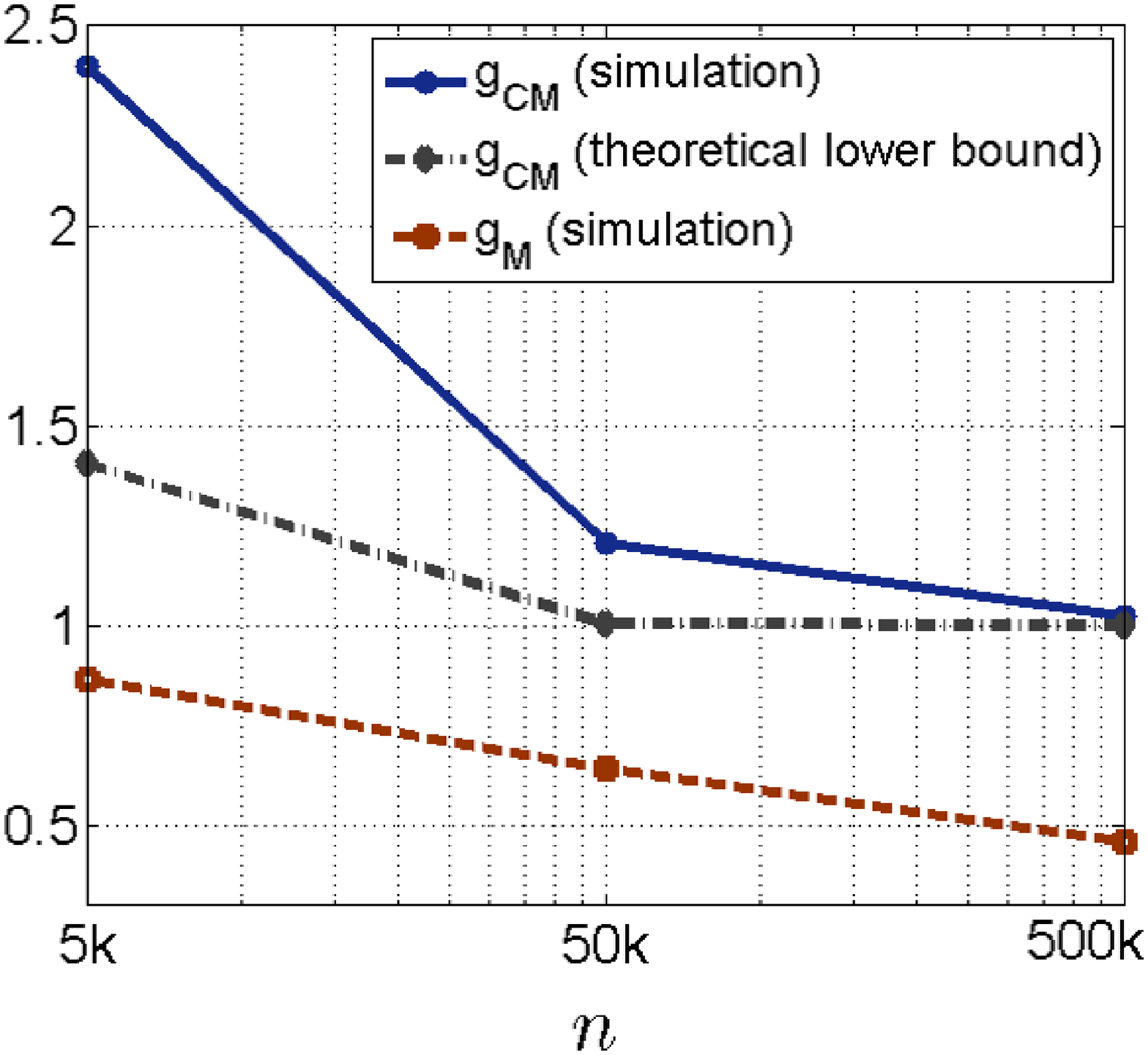}
  \vspace{-0.1in}
  \caption{Theoretical and simulation results for $g_\text{M}$ and $g_\text{CM}$.}
  \vspace{-0.1in}
  \label{fig:sim_theory}
  \end{minipage}
\vspace{-0.15in}
\end{figure*}

\subsection{{\UCM}: Case $\mc{K}\geq 2$}
\label{subsec:gain_UcompCM}
Thus far, we demonstrated in Proposition~\ref{lem:no_cluster} that the crude memorization in the memory unit is not beneficial when a compound source is present. This necessitates to first appropriately \emph{cluster} the sequences in the memory. Then, based on the criteria as to which cluster the new sequence $x^n$ belongs to, we utilize the corresponding memorized context for the compression.
In the following, we analyze the problem for the source-defined clustering (defined in Sec.~\ref{sec:gain_setup}). In this clustering, we assume both $S$ and $R$ (in Fig.~\ref{fig:single_hop}) can exactly know the index $i \in [\mc{K}]$ and hence all the sequences that belong to the same source $\theta^{(i)}$  in the compound source are assigned to the same cluster (for all $i \in [\mc{K}]$). Further, we assume that $S$ can exactly \emph{classify} the new sequence $x^n$ to the cluster with parameter $\theta^{(Z_T)}$.
In Sec.~\ref{sec:cluster}, however, we will relax these assumptions and study the impact of  clustering in practice.

Let $H(\mb{p}) = -\sum_{i=1}^\mc{K} p_i \log(p_i)$ be the entropy of the source model. The following proposition quantifies the achievable redundancy and the memorization gain of UcompCM.
\begin{theorem}
Let $\theta^{(i)}$ ($i \in [\mc{K}]$) follow Jeffreys' prior. Then, the memorization gain of UcompCM is lower bounded by
\begin{equation}
g_\TCM(n,m,\phi,\epsilon,\mb{p} )
\hspace{-.02in}\geq\hspace{-0.03in} 1 \hspace{-.01in}+ \frac{\bar{R}_n + \log(\epsilon )  - \hat{R}_{2}(n,m)}{H_n(\phi) + \hat{R}_{2}(n,m)}  + O\hspace{-0.03in}\left(\frac{1}{n\sqrt{m}} \right)\hspace{-0.05in},\nonumber
\end{equation}
where
$\hat{R}_{2}(n,m) \triangleq \frac{d}{2}\log\left( 1+ \frac{n}{p_{Z_T} m}\right) + 3 + H(\mb{p})$.
\label{lem:UcompCM}
\end{theorem}

\section{Clustering for Memory-Assisted Compression}
\label{sec:cluster}
In this section, we try to answer the main question in the memory-assisted compression setup we introduced: ``How do we utilize the available memory to better compress a sequence generated by a compound source?''
It is obvious that the performance of conventional universal compression schemes (those without memory) cannot be improved by clustering of the compound source as $x^n$ is encoded without regard to $\mb{y}$. However, because of a compound source, clustering is necessary to effectively utilize the memory in the proposed memory-assisted compression. Within this framework, we identify two interrelated problems:
1) How do we perform clustering of the memorized data to improve the performance of memory-assisted compression?
2) Given a set of clustered memory, how do we classify an incoming new sequence  into one of the clusters in the memory using which the performance of memory-assisted compression is maximized?
This relaxes the assumption of knowing $\mb{Z}$ by the encoder and the decoder in the analysis of Sec.~\ref{sec:results}.

 As one approach, it is natural to adapt a clustering algorithm, among the many, that has the codelength minimization as its principle criterion.
 Thus, the goal of the clustering is to group the sequences in the memory such that the total length of all the encoded sequences is minimized (i.e., the sequences are grouped such that they are compressed well together).
 We employ a Minimum Description Length (MDL)~\cite{Barron_MDL} approach suggested by~\cite{Kontkanen2003}. The MDL model selection approach is based on finding shortest description length of a given sequence relative to a model class.
We do not have a proof that the MDL clustering is necessarily optimal for our goal (for all sequence lengths and memory sizes). However, as we will see in Sec.~\ref{sec:simulation}, for the cases of interest where the length of memory is larger than the length of the new sequence, the MDL clustering demonstrates a very good performance close to that of assuming to know $\mb{Z}$ (in source-defined clustering).

Now, we would like to find a proper class for a new sequence $x^n$ generated by one of the $\mc{K}$ sources.
Given a set of $T$ sequences taken from $\mc{K}$ different sources, we assume those $T$ sequences have already been clustered into $\mc{K}$ clusters $C_1, \ldots, C_\mc{K}$.
 Then, the classification algorithm for $x^n$ is as follows.
We include the sequence $x^n$ in each cluster one at a time and find the total description length of all sequences in the $\mc{K}$ clusters. Then, we label $x^n$ with the cluster whose resulting total description length is the minimum.

Next, we describe as to how we cluster the $T$ sequences in memory. A good clustering is such that it allows efficient compression of the whole data set. Equivalently, the sequences that are clustered together should also compress well together. We start by an initial clustering of the data set in the memory. Through experiments, we observed that this initial clustering has a considerable impact on the convergence of the clustering algorithm which is in accordance with the observation in~\cite{Kontkanen2003}. We found that an initial clustering based on the estimated entropy of the sequences greatly reduces the number of iterations till convergence.
The clustering is done iteratively by surfing the data set and moving a sequence from cluster $i$ to cluster $j$ if this swapping results in a shorter total description length of the whole data set.

\vspace{-0.1in}
\section{Simulation Results and Conclusion}
\vspace{-0.05in}
\label{sec:simulation}

In this section, we characterize the performance of the proposed memorization scheme through computer simulations. In order to illustrate the importance of clustering for efficient use of memory, we have evaluated the memory-assisted compression gain (over the performance of the conventional universal compression) for three cases: $g_\TM$, $g_\text{CM}$, and $g_\text{MDL}$. Note that $g_\text{MDL}$ is defined as the gain of memorization with MDL clustering in Sec.~\ref{sec:cluster}.
We demonstrate the significance of the memorization through an example, where we again consider $\mc{K}$ first-order Markov sources with alphabet size $k=256$, source entropy $\frac{H_n(\phi)}{n} =1$ bit per byte, and $\epsilon = 0.05$.

\begin{table}
\caption{Memory-assisted compression gain under MDL clustering}
\label{tab:g_mdl}
\centering
\begin{tabular}{||l|c|c||}
\hline
	~					&$n=10$KB 	&$n=100$KB	\\
\hline
	$g_{\text{MDL}}$ ($m=1$MB)								&1.7308		&1.0372 \\
\hline
	$g_{\text{MDL}}$ ($m=10$MB)							&1.8862		&1.0939	\\
\hline
	$g_{\text{MDL}}/g_{\text{CM}}$ ($m=1$MB)								&0.944		&0.993\\
			
\hline
	$g_{\text{MDL}}/g_{\text{CM}}$	($m=10$MB)							&0.939		&0.998\\
\hline
\end{tabular}
\vspace{-.2in}
\end{table}

 Fig.~\ref{fig:M1_256_gain_n} considers the single source case (i.e., $\mc{K} = 1$). The lower bound on the memorization gain is demonstrated as a function of the sequence length $n$ for different values of the memory size $\MM$.
As can be seen, significant improvement in the compression  may be achieved using memorization.
 As demonstrated in Fig.~\ref{fig:M1_256_gain_n}, the memorization gain for a memory of length $m=8$MB is very close to $g(n,\infty,\phi,0.05, 1)$, and hence, increasing the memory size beyond $8$MB does not result in the substantial increase of the memorization gain.
We observe that more than $50 \%$ improvement is achieved in the compression performance of a sequence of length $n=128$kB with a memory of  $m=8$MB.
On the other hand, as $n\to\infty$, the memorization gain becomes negligible as expected.

For the rest of the experiments, we fixed the length of the sequences generated by the source, i.e., $n_j =n$ for all $j$. We used $\mc{K} = 10$ with uniform distribution, i.e., $p_i = \frac{1}{10}$ for $i \in [\mc{K}]$. Further, we performed compression using Context Tree Weighting (CTW)~\cite{CTW95} and averaged the simulation results over multiple runs of the experiment.
Fig.~\ref{fig:g_pc} depicts $g_\text{CM}$. As can be seen, joint memorization and clustering achieves up to 6-fold improvement over the traditional universal compression.
Fig.~\ref{fig:sim_theory} depicts the experimental $g_\text{CM}$ using CTW, the theoretical lower bound on $g_\text{CM}$ derived in Sec.~\ref{sec:results}, and the experimental $g_\text{M}$ for memory $m=10$MB.
As we expected, with no clustering, the memory-assisted compression may result in a worse compression rate than compression with no memory validating our theoretical result in Sec.~\ref{subsec:gain_UcompM}. Finally, the experimental results of memory-assisted compression gain $g_\text{MDL}$ under MDL clustering, summarized in Table~\ref{tab:g_mdl}, show that $g_\text{MDL}$ is close to $g_\TCM$, demonstrating the effectiveness of MDL clustering for compression.

In conclusion, this paper demonstrated that memorization (i.e., learning the source statistics) can lead to a fundamental performance improvement over the traditional universal compression.
This was presented for both single and compound sources. We derived theoretical results on the achievable gains of memory-assisted source coding for a compound (mixture) source and argued that clustering is necessary to obtain memorization gain for compound sources.
We also presented a fast MDL clustering algorithm tailored for the compression problem at hand and demonstrated its effectiveness for memory-assisted compression of finite-length sequences.

\ifCLASSOPTIONcaptionsoff
  \newpage
\fi



%


\bibliographystyle{IEEEtran}
\bibliography{compress}

\end{document}